\documentclass[prl,twocolumn]{revtex4-1}%
\usepackage{amsmath}
\usepackage{amsfonts}
\usepackage{amssymb}
\usepackage{graphicx}%
\begin{document}

\title{Approaches to Three-Dimensional Transformation Optical Media Using Quasi-Conformal Coordinate Transformations}
\author{N.~I. Landy}
\affiliation{Department of Electrical and Computer Engineering, Duke
University, Durham, NC 27708 USA.}
\author{N. Kundtz}
\affiliation{Department of Electrical and Computer Engineering, Duke
University, Durham, NC 27708 USA.}
\author{D. R. Smith}
\affiliation{Department of Electrical and Computer Engineering, Duke
University, Durham, NC 27708 USA.}

\begin{abstract}
We introduce an approach to the design of three-dimensional transformation optical (TO) media based on a generalized quasi-conformal mapping approach. The generalized quasi-conformal TO (QCTO) approach enables the design of media that can, in principle, be broadband and low-loss, while controlling the propagation of waves with arbitrary angles of incidence and polarization. We illustrate the method in the design of a three-dimensional ``carpet'' ground plane cloak and of a flattened Luneburg lens. Ray-trace studies provide a confirmation of the performance of the QCTO media, while also revealing the limited performance of index-only versions of these devices.
\end{abstract}
\maketitle

Transformation optics (TO) is a unique tool for the design of complex electromagnetic media\cite{Pendry06232006}. TO makes use of the form-invariance of Maxwell's equations to mimic spatial transformations using distributions of inhomogeneous and anisotropic material constitutive parameters. TO has inspired many exotic devices, one of the most compelling of which being the electromagnetic ``invisibility'' cloak. However, the use of TO as a design methodology typically comes at high cost; the media derived from coordinate transformations generally involve spatial gradients in all nine components of the epsilon and mu tensors. Though it is possible to find a basis that diagonalizes these tensors, the diagonal basis will generally be a function of position for all but the most simplistic and symmetric designs.  Moreover, the required response is generally outside of the range of natural materials.

Electromagnetic metamaterials (MMs) are used to access the extreme material parameters required by TO media. MMs, for example, were used to demonstrate a negative index of refraction\cite{PhysRevLett.84.4184,R.A. Shelby04062001} and electromagnetic cloaking \cite{D.Schurig11102006}. However, the performance of these initial MM constructs was limited by a combination of narrow bandwidth and relatively large absorption. The typical limitation for MM designs has been the requirement of constitutive parameters that have a large range of values for both permittivity and permeability. The implementation of artificial magnetism, in particular, requires resonant inclusions that are inherently lossy and dispersive, leading to absorption and reduced bandwidth. More recently, the development of coordinate transformation methods in optical design approaches has significantly advanced MM complexity: independent magnetic and electric responses are required in all directions for general TO designs, yet most MM elements provide a controlled response in one or two directions. Were one to attempt to control all of the tensor elements of a MM simultaneously, multiple MM elements would need to be either co-located or closely positioned, introducing very complicated magneto-electric coupling difficult to control or even quantify using current MM retrieval techniques \cite{PhysRevE.81.036605}.

Fortunately, the enormous degree of flexibility available in coordinate transformations can alleviate many of the complexities and limitations of TO media. For a given TO device that performs some function, there is typically little concern over the physical distribution of fields within the device itself; rather, it is the distribution of fields external to the TO medium that is of consequence. Therefore, the freedom exists to choose any of an infinite number of possible coordinate transformations, subject to the boundary conditions, to achieve the same functionality while simultaneously optimizing certain design criteria. In 2008, Li and Pendry introduced the concept of the quasi-conformal map (QCM) to TO\cite{liCarpet}. It may be shown that the QCM minimizes anisotropy across a two-dimensional transformation. Applying QCM optimization, Li and Pendry conjectured that certain classes of TO media could be realized using dielectric-only materials. The approach was applied to the design of a medium that would mask a deformation introduced into an otherwise flat, perfectly-conducting (PEC) plane. The QCM technique has subsequently been applied in several experiments, and has also been used to demonstrate that substantial improvements can be achieved in the redesign of more conventional imaging devices\cite{kundtzLuneburg}.

While the advantages of dielectric-only QCM-derived media are extremely important from an implementation point-of-view, such media can only control waves propagating in two-dimensions. In this paper we show that the QCM concept can be readily generalized for the design of media that can control wave propagation in three-dimensions (3D); however, though greatly simplified, 3D QCM media are inherently anisotropic, due to the requirement of magnetic response in at least one direction. We illustrate the performance of generalized 3D QCM media with ray-traces of a carpet cloak and of a flattened Luneburg lens, comparing with 2D QCM media. The ray traces enable us to quantify the effectiveness of the methods using the standard metrics of optical lens design. Finally, we introduce a method to mitigate the effect of the approximations that are made with QCM and QCM-like mappings.

Consider an arbitrary mapping in of the form $x^{i'}=x^{i'}(x^i)$.
The unprimed coordinates represent the virtual domain (subscript ``v''), and the primed coordinates represent the physical domain (subscript ``p''). The virtual domain is the domain as perceived by electric and magnetic fields. For cloaks, this domain is typically vacuum. The physical domain is the coordinate system that exists in physical space. The TO-derived material tensors ($\bar{\bar{\epsilon}}=\bar{\bar{\mu}}$) for this transformation are
\begin{equation}
\epsilon^{i'j'}=\mathrm{Det}(\Lambda)^{-1}\Lambda^{i'}_{i}\Lambda^{j'}_{j}\epsilon^{ij}
\end{equation}
where $\Lambda^{i'}_{i}=\partial{x^{i'}}/\partial{x^{i}}$ is the Jacobian of the transformation.
Suppose that we wished to find a transformation that required no magnetic coupling. Since $\epsilon^{i'j'}=\mu^{i'j'}$, the magnetic response can only be eliminated by the identity transformation. However, if a transformation could be found such that $\epsilon^{i'j'}=\epsilon^{i'j'}\delta_{i'j'}$ and $\epsilon_{x}=\epsilon_{y}=1$, then for waves polarized such that the electric field is constrained to lie in the $\hat{z}$ direction, only $\epsilon_{z}$ is relevant and the transformation could be realized with a dielectric-only medium. A conformal transformation accomplishes exactly this goal. However, a coordinate mapping between a physical and a virtual domain can only be conformal between two domains that share the same conformal module, or $M = m_v/m_p=1$. If $M\neq1$, then the transformation is quasiconformal, and the in-plane tensor components of the material parameters will no longer be equal. Additionally, orthogonality is maintained at the boundary of the mapping via Neumann (``slipping'') boundary conditions (BCs). These BCs will generally introduce discontinuities at the boundaries that manifest as refractive aberrations and reflections.

\begin{figure}\label{MapInvariant}
\includegraphics[width=3.33in]{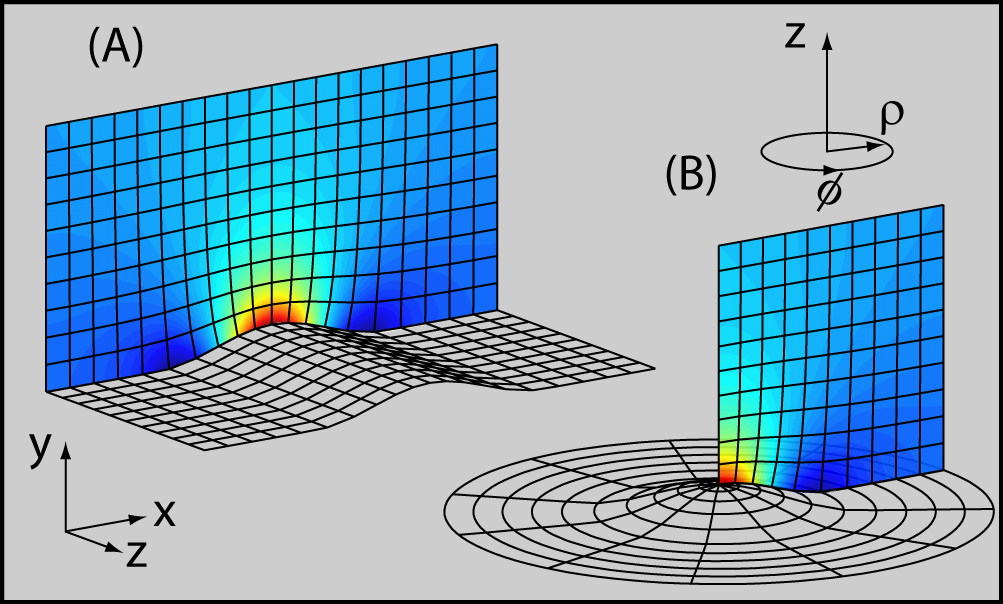}
\caption{(Color Online) 3D carpet cloaks created using 2D quasi-conformal transformations. In (A) the mapping is performed in x and y, and z is invariant. In (B), the mapping is performed in $\rho$ and $z$, while $\phi$ is invariant.}
\end{figure}

It is important to note that QC mappings are two dimensional (e. g. x and y) with the third dimension (z) $invariant$ under the transformation. Therefore, the QCTO prescription is valid as long as the geometry is also invariant along $z$. An example of a structure for which a QC transform solution can be found is the carpet cloak, shown in figure \ref{MapInvariant}(a). For a wave polarized along the z-direction and propagating in the xy plane, a dielectric only implementation can be applied; however, the full TO solution requires a magnetic response equal to the electric response in every direction. Therefore, we require that $\mu_{z}=\epsilon_{z}$ for the full TO medium.

The TO approach can be fruitfully applied to systems such as waveguides\cite{landyBends} and transmission lines. However, many systems of great interest are invariant $azimuthally$; for such systems, it is then natural to seek an analog QCM mapping in cylindrical coordinates. Assuming $M\approx1$, we calculate the QCM $x_p(x_v,y_v)$ and $y_p(x_v,y_v)$, and then make the substitutions $x_{v,p}\rightarrow\rho_{v,p}$ and $y_{v,p}\rightarrow{z_{v,p}}$. We then determine the material transformation in cylindrical coordinates\cite{Pendry06232006,1367-2630-10-11-115034,kundtzThesis}. Figure \ref{MapInvariant}(b) shows an example of such an azimuthally invariant geometry and map. Remembering that $\bar{\bar{\mu}}=\bar{\bar{\epsilon}}$, we obtain the following material parameters:
\begin{equation}\label{full}
\epsilon_\phi = \alpha/\beta\text{, and }\epsilon_\rho\approx\epsilon_z\approx\beta
\end{equation}
where $\alpha = |\Lambda|^{-1}$ and $\beta=\rho_v/\rho_p$. The factor $\beta$ arises from the fact that the differential volume element in cylindrical coordinates is a function of $\rho$. The transformed material parameters must compensate for the dilation of space between the virtual and physical coordinates; however, the transformed medium remains approximately uniaxial. A further simplification can be made if we are mostly interested in a structure that controls ray propagation in the eikonal limit, where only the anisotropic refractive index is of consequence\cite{Roberts:09}. In the eikonal limit, we can thus make the substitutions
\begin{equation}\label{reduced}
\epsilon'_\rho=\epsilon'_z=\beta^2\text{, }\epsilon'_\phi = \alpha\text{, }\mu'_\rho=\mu'_\phi=1\text{, and }
\mu'_\phi = \alpha/\beta^2
\end{equation}
to achieve an equivalent medium. Since most magnetically-coupled MMs provide a magnetic response in only one direction, the substitutions of (\ref{reduced}) ease implementation by requiring only one component of $\mu$ to deviate from unity. The eikonal approximation may or may not be appropriate depending on the intended operation. Note that neither (\ref{full}) nor (\ref{reduced}) are isotropic material specifications. Therefore, a purely dielectric implementation of this mapping cannot control waves with arbitrary incidence and polarization, even in the eikonal limit.

We can understand the need for anisotropy by considering the local dispersion relation for the system, which for our transformation is given by:
\begin{equation}\label{dispersion}
{k_\phi^2}/\beta^2+k_\rho^2/\alpha+k_z^2/\alpha-1=0.
\end{equation}
where the wavenumber $\mathbf{k}$ has been normalized to the free-space value $k_0$. This form of the dispersion relation is derivable directly from Maxwell's equations in a homogeneous, index-matched ($\bar{\bar{\epsilon}}=\bar{\bar{\mu}}$) medium with cylindrical anisotropy. This dispersion relation can be achieved by an isotropic dielectric $\epsilon = \alpha$ only if $k_\phi=0$. The cylindrical symmetry of the system ensures that $k_\phi=0$ along the ray's entire trajectory, and the ray would therefore be properly transformed. This means that only rays in planes of constant $\phi$ are properly directed (meridional rays) for a dielectric only medium. We note that the dispersion relation is actually fourth-order in $k$, but we have suppressed a power of two because the two modes are degenerate for impedance-matched media\cite{Schurig:06}. Our substitution in (\ref{reduced}) preserves this degeneracy, and therefore our prescription remains singly refracting.
\begin{figure}\label{CarpetComp}
\includegraphics[width=3.33in]{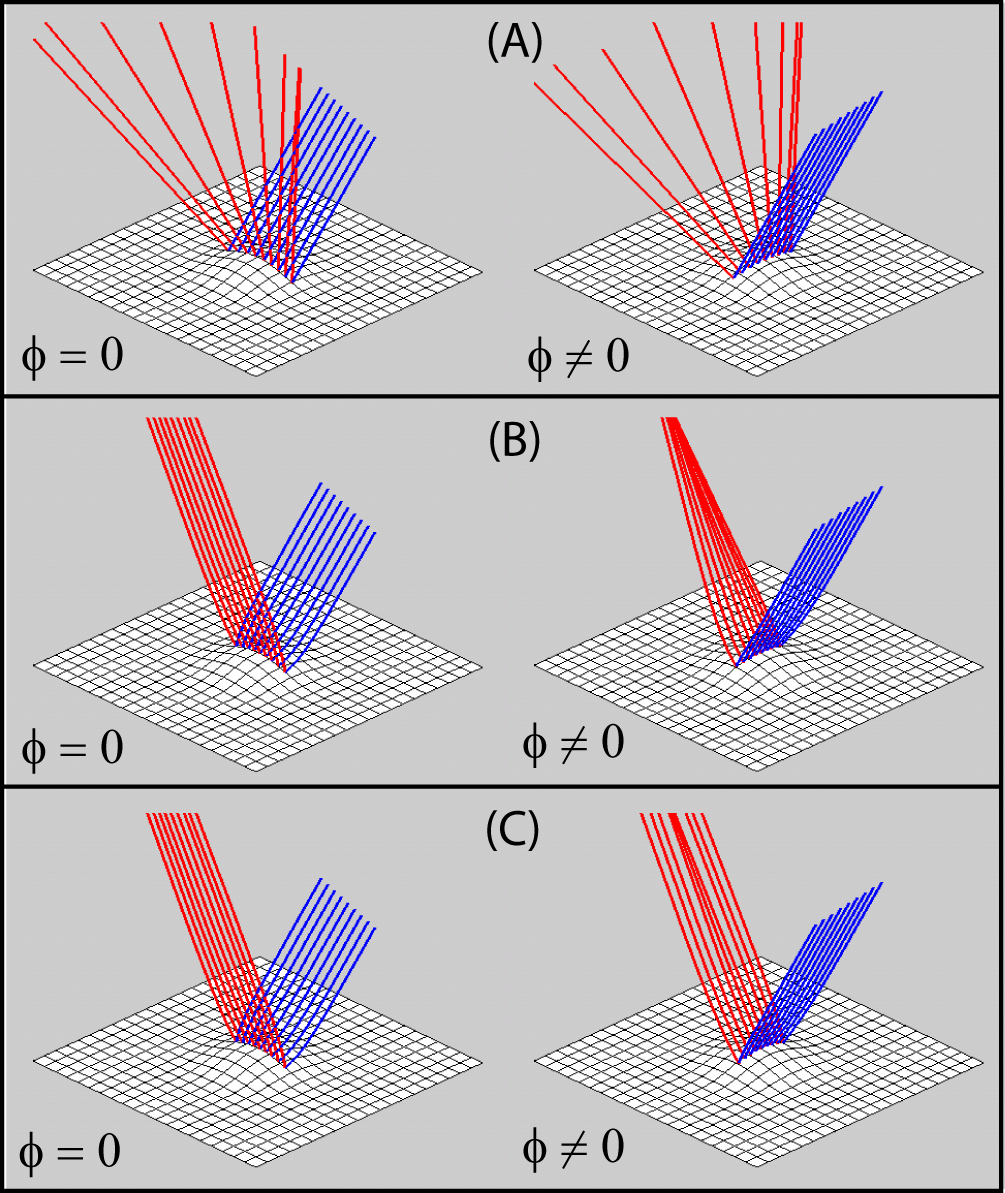}
\caption{(Color Online) Ray-tracing comparisons of (A) The bare PEC scatterer, (B) the index-only revolved carpet cloak and (C) the properly transformed (anisotropic) carpet cloak. The angle of incidence is $30^{\circ}$. The plots on the left consist of incident rays in the plane containing the optical axis. The plots on the right consist of rays off the optical axis. The incident rays are blue and the reflected rays are red.}
\end{figure}

We can demonstrate the limitations of dielectric-only implementations, as well as the validity of 3D QCM media via ray tracing\cite{Schurig:06}. We begin by studying the 3D carpet-cloak \cite{liCarpet,R.Liu01162009,ISI:000267204600015,ISI:000268935200019,TolgaErgin04162010,maCloak,PhysRevLett.104.233903,PhysRevLett.104.253903}, designed to effectively flatten a perturbation of the form $z=0.1\times\cos(\pi\rho)^2$ in the region $\rho<0.5$. Ray-tracing analyses of both the dielectric-only and anisotropic implementations are shown in figure \ref{CarpetComp}. A ray trace from the uncloaked perturbation is shown for comparison. In both cloaks, Rays incident such that $k_\phi=0$ are properly reflected at the specular angle. However, when $k_\phi \neq 0$ The dielectric-only implementation does not properly redirect the rays. Rather, the rays appear to be slightly focused by the index distribution. Rays that do not lie on the optical axis experience a gradient transverse to their initial trajectories that gradually bends the rays around the optical axis. Therefore, rays that were initially parallel come to lie on intersecting trajectories after the exit the device. Nevertheless, the index-only cloak clearly reduces scattering away from the specular direction, possibly masking the distortions caused by the cloak in an experimental setting.

The anisotropic transformation described by (\ref{full}), properly cloaks the bump. The reflected rays exit the cloak as if they had reflected specularly from a flat ground plane. The rays do refract slightly at the boundaries of the mapping as the material parameters differ slightly from unity. Refraction at the mapping boundary is included in our ray-tracing analysis, but its effect appears to be negligible.
\begin{figure}\label{LensMatParam}
\includegraphics[width=3.33in]{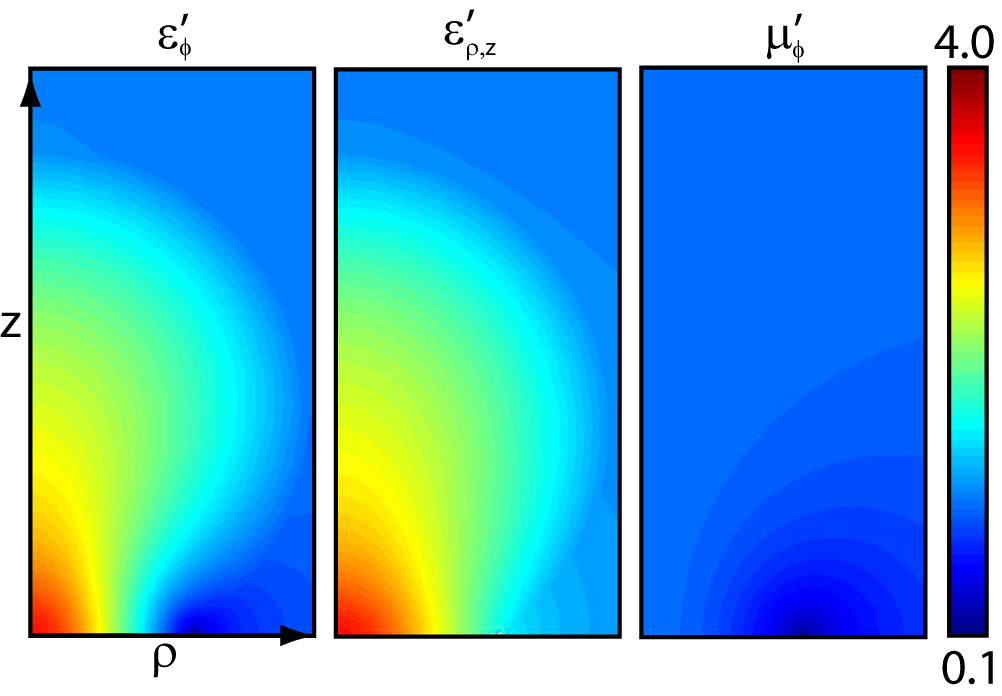}
\caption{(Color Online) Material parameters for the 3D QCM flattened Luneburg Lens. Only the tensor components that differ from unity are shown.}
\end{figure}
\begin{figure}\label{LensComp}
\includegraphics[width=3.33in]{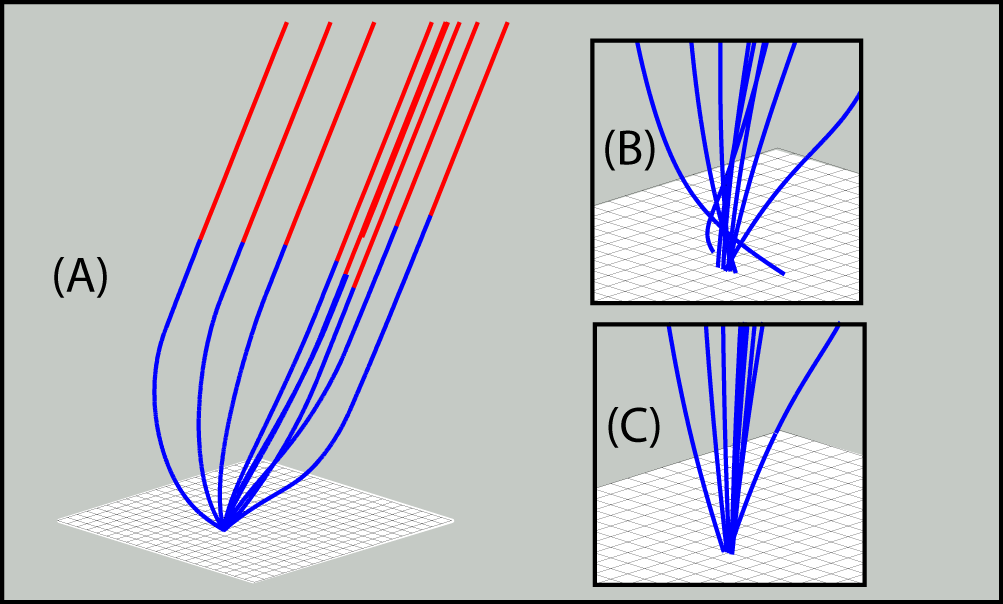}
\caption{(Color Online) (A)Ray-traces of collimated rays incident on a flattened Luneburg lens. (B) is a close-up of the index-only revolved Luneburg lens and (C) the properly transformed (anisotropic) lens. The angle of incidence is $30^{\circ}$. The rays are red outside of the mapped region and blue inside.}
\end{figure}

The QCM technique holds great potential for its ability to modify and improve conventional optical devices. The Luneburg lens, for example, is a rotationally-symmetric, gradient-index medium that perfectly images two concentric spherical surfaces on to one another\cite{luneburg}. If one of these surfaces is taken to infinity, then parallel rays are focused to the finite surface. The permittivity distribution for the Luneburg is given by\cite{luneburg} $\epsilon_l = 2-(r/a)^2$, where $a$ is the radius of the lens.

Despite its powerful imaging capabilities, the Luneburg is seldom used since the spherical focal surface is incompatible with conventional planar detector arrays. It was proposed that TO could be used to flatten the focal surface and retain the same imaging performance\cite{1367-2630-10-11-115034}. The flattened Luneburg was subsequently realized by using the QCM method in 2D\cite{kundtzLuneburg}. A 3D version of the lens can be obtained, following the procedure given above, by a medium whose non-unity parameters are given by:
\begin{equation}\label{reducedLuneburg}
\epsilon'_\rho=\epsilon'_z=\epsilon_l\beta^2\text{, }\epsilon'_\phi = \epsilon_l\alpha\text{, and }\mu'_\phi = \alpha/\beta^2
\end{equation}
where it is understood that $\epsilon_l=\epsilon_l(\rho_v,z_v)$. We consider a lens that has been flattened to provide a field-of-view (FOV) of $90^\circ$. The reduced material parameters for this QCM are shown in figure \ref{LensMatParam}. We also examined an index-only implementation by making the assignment $\epsilon_\phi \rightarrow \epsilon$.

\begin{figure}\label{SpotComp}
\includegraphics{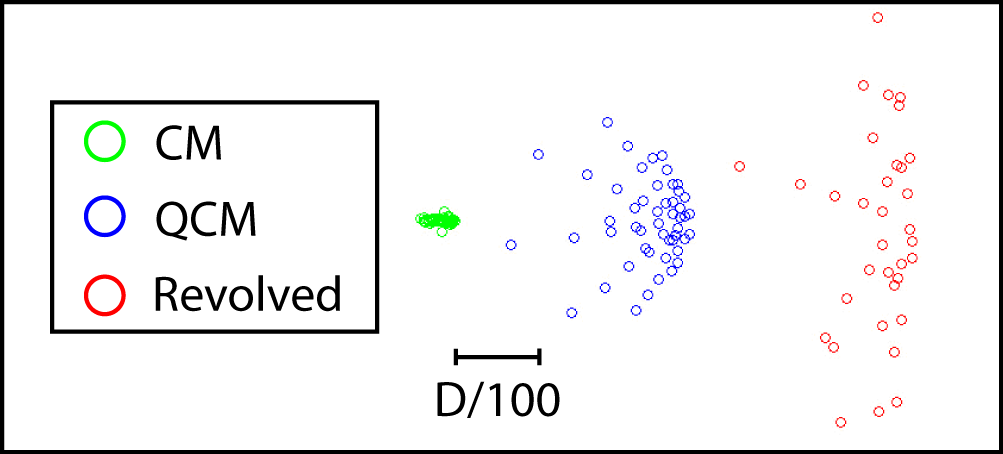}
\caption{(Color Online) Ray-intercepts of the focal planes for three flattened Luneburg designs. 50 collimated rays illuminating 90$\%$ of the aperture are traced through the lens. The angle of incidence is $30^{\circ}$. A scale bar is included to show the spot size in terms of the unflattened lens diameter, D.}
\end{figure}

Ray-traces depicting the focal properties of both lenses are shown in figure \ref{LensComp}. The collimated rays are incident at an angle of 30$^\circ$. The anisotropic lens in figure \ref{LensComp}(b) shows good focusing at 30$^\circ$, whereas the index-only lens in \ref{LensComp}(c) shows severe off-normal astigmatism. This is clearly seen in a plot of the ray intercepts at the flattened focal plane (figure \ref{SpotComp}. Rays parallel (but not on) the plane containing the optical axis are over-steered such that they cross the plane containing the optical axis before they intersect the bottom of the lens.
Even though the anisotropic lens offers better performance than its index-only counterpart, there are still aberrations present due to the uniaxial approximation in \ref{reduced}. This approximation  effectively stretches the virtual domain in one direction\cite{PhysRevLett.104.233903}, such that the Luneburg lens is no longer spherical in the virtual domain.

For a FOV of 180$^\circ$, the QCM must be used for boundary correspondence on three sides of the map. However, for a restricted FOV (such as 90$^\circ$), we have the freedom to change the dimensions of the virtual domain so that $M=1$ and the QCM reduces to the CM\cite{landyBends}. The material parameters of the CM are nearly identical to those shown in figure \ref{LensMatParam} by eye, but aberrations in \ref{SpotComp} are visibly reduced. The only remaining aberrations are caused by slight refraction at the top boundary of the map. Since there is a small gradient in the material parameters across the top boundary, collimated rays are refracted at different angles upon entering the transformed domain. Since these rays are no longer parallel, they are focused to different points on the focal plane.  These aberrations are only reduced by extending the boundaries of the map, thereby increasing the size of the device. By changing the module in this fashion, we have also increased the material discontinuities on the sides of the map. This restricts the FOV of the lens.

We have shown that the QCM technique can be extended to the design of rotationally symmetric TO devices that can be realized using simplified, uniaxial media. The Luneburg mapping illustrates several important features that are common to the 3D QCM solutions and could allow for broadband operation.  The dielectric response is mostly positive and moderate everywhere, with $\epsilon'_{\rho,z}$ ranging from 1 to approximately 3.27 and $\epsilon'_{\phi}$ ranging from approximately 0.22 to 3.35. However, throughout most of the transformed region,  $\epsilon'_{\phi}>1$ and we expect that we can retain most of the FOV of the lens while neglecting the portion with values falling below the vacuum permittivity. Though magnetic response is required, $\mu'_{\phi}$ ranges from approximately 0.15 to one and can thus potentially be broadband. Broadband diamagnetism has been previously suggested, with one possible implementation consisting of stacked metal plates oriented to provide a magnetic response in the azimuthal direction\cite{0953-8984-19-7-076208}. The tremendous simplification in material properties afforded by the 3D QCM approach increases the viability of many potential TO devices.

This work was supported by the Army Research Office through a Multidisciplinary University Research Initiative (MURI), Grant No. W911NF-09-1-0539. NBK also acknowledges support from the IC postdoctoral fellowship program.

\bibliography{references}

\end{document}